\documentclass[%
 reprint,
 a4paper,
superscriptaddress,
unsortedaddress,
 amsmath,amssymb,
prb,
floatfix,
]{revtex4-2}

\usepackage[utf8]{inputenc}
\usepackage[OT2,T1]{fontenc}
\usepackage[english]{babel}
\usepackage{cancel}
\usepackage{graphics}
\usepackage{amsmath,amsfonts,bm}
\usepackage{amssymb}
\usepackage{bigints}
\usepackage{color}
\usepackage{dsfont}
\usepackage{comment,soul}

\usepackage[oldvoltagedirection]{circuitikz}
\usepackage{tikz}
\usetikzlibrary{patterns,arrows,shapes,arrows,intersections,decorations}

\usepackage{graphicx}
\usepackage{dcolumn}
\usepackage{bm}

\renewcommand{\vec}[1]{\mathbf{#1}}

\begin{document}


\title{Hydrodynamic electrons in Graphene: a viscous boundary-layer description.
}

\author{Pedro Cosme}
\email{pacsilva@ipfn.tecnico.ulisboa.pt}
\affiliation{Instituto de Plasmas e Fusão Nuclear}
\affiliation{Instituto Superior Técnico 1049-001 Lisboa, Portugal
}%

\author{João S. Santos}
\affiliation{Instituto Superior Técnico 1049-001 Lisboa, Portugal}

\author{Hugo Terças}
\affiliation{Instituto de Plasmas e Fusão Nuclear}
\affiliation{Instituto Superior Técnico 1049-001 Lisboa, Portugal
}
%


\date{\today}

\begin{abstract}

In this paper we dwell over the study of the boundary layer problem in a hydrodynamical description of the electrons in gated graphene. It has been verified experimentally that this fluid can display non-Poiseuille like flow as reproduced in our numerical simulation. In fact, the velocity profile displays a maximum value close to the boundary and then decreases as it approaches the bulk of the graphene layer. This work aims to present a satisfactory theoretical description of the boundary layer problem in graphene. We found that by using the fluid equations and following a method similar to that for deriving Blasius' equation, a non-linear model can be obtained whose solutions display the maximum values of velocity near the edges of the graphene layer. We argue that such a non-monotonic model and behaviour can shed some light on the subject of non-topological edge currents in graphene.

\end{abstract}

\maketitle


\section{Introduction}

The study of the collective state of carriers, be it electrons or holes, in graphene is a difficult task to endeavour, particularly in a fully quantum description with few works \cite{Figueiredo2020WignerPlasmasb} dedicated to that goal. Instead, a growing interest in semi-classical descriptions can be found. In particular, several hydrodynamics descriptions of the mean field quantities have been brought forward in the late years \cite{Cosme2020,Lucas2018,Narozhny2017,Svintsov2013,Barletti2014}, with some experimental works also taking the first steps \cite{Sulpizio2019,Ku2020a}. 

For such hydrodynamical models, a central point of discussion is the nature of the flow, that is, what kind of velocity profile is foreseeable under a specific set of conditions, as with regular hydrodynamics this is ought to be primarily determined by the viscosity of the fluid. 
Nonetheless, to calculate the viscosity of such system from first principles is a complex task, and even if the results for undoped graphene are relatively consensual \cite{Principi2016BulkSheet,Narozhny2019MagnetohydrodynamicsViscositiesb} the same does not happen for the doped scenario, seemingly the most relevant situation for technological applications.  Moreover, the boundary conditions for the Fermi liquid at the edges of a graphene layer are still not completely understood and the naive approach of a no-slip condition might not be the most accurate one\cite{Kiselev2019BoundaryFlow,Danz2020VorticityGrapheneb}.  

Some authors have committed their analysis to a highly viscous regime, leading to a Poiseuille flow \cite{Sulpizio2019,Polini2019}. However, for such a parabolic flow to develop there are two conditions required; namely, that the flow is quasi-stationary, i.e. it has a small Womersley number \cite{Moessner2018a}, and that the transverse dimension of the system is smaller than the boundary layer\cite{Schlichting2017Boundary-layerEdition}, i.e. in a narrow channel. These condition might not be met in some conditions, either when dealing with regimes with high frequency or wide channels, for instance in the case of graphene nano-antennae operating at THz \cite{Cosme2020,Bianco2020Antenna-coupledArray,Dash2020}. Thus, the assessment of the boundary layer thickness for the electronic flow is of central importance for study of the hydrodynamic models of conduction in graphene. 

 In this work we present a description of the boundary layer for graphene electrons, in the degenerate limit. We start by establishing the hydrodynamic model for the conducting electrons in gated graphene \cite{Cosme2021} and, in the following sections, proceed to derive a nonlinear model for the velocity profile across the channel, comparing it with the simulations from which we also extract scaling laws for the boundary layer thickness.

\section{Electron Fluid in Graphene}


In the highly degenerate regime, the electrons in graphene are all described by a hydrodynamic model resorting to the Drude mass \cite{Lucas2018,Svintsov2013}
\begin{equation}
    m^\star=\frac{\hbar k_F}{v_F}=\frac{\hbar\sqrt{\pi n}}{v_F},
    \label{eq:DrudeMass}
\end{equation}
where $\hbar k_F$ is the Fermi momentum and $n$ is the carrier number density. It is important to note that, since the electronic fluid is compressible, the effective mass is not a conserved quantity, contrary to customary fluids.

From a kinetic description one can derive \cite{Cosme2021} the conservation laws for number of particles, momentum and energy as

\begin{subequations}
\begin{equation}
\frac{\partial n}{\partial t}+\bm{\nabla}\!\cdot\!\frac{\vec{p}}{m^\star}=0, \label{eq:continuity}   
\end{equation}
\begin{equation}
\frac{\partial \vec{p}}{\partial t}+\bm{\nabla}\!\cdot\!\left( \frac{ \vec{p}\otimes \vec{p}}{nm^\star} + \mathbb{P}\right)+en\bm{\nabla}\phi=0.\label{eq:consMom}
\end{equation}
\begin{equation}
      \frac{\partial \mathcal{E}}{\partial t}+\bm{\nabla}\cdot\left(\frac{3}{2}\mathcal{E}\vec{v}\right)+e\frac{\vec{p}}{m^\star}\cdot\bm{\nabla}\phi=0, 
      \label{eq:consEnergy}
\end{equation}\label{eq:model}\end{subequations}
respectively. Where $\vec{p}=nm^\star\vec{v}$ is the fluid two-dimensional momentum density, $\mathcal{E}$ the energy density, $\mathbb{P}$ the pressure stress tensor, and $\phi$ the electrical potential. In the present work however, we will not make use of the energy conservation equation \eqref{eq:consEnergy} since the continuity \eqref{eq:continuity} and Cauchy \eqref{eq:consMom} equations will form a closed set of equations by themselves as will become evident. 
 
The two-dimensional pressure term in \eqref{eq:consMom} can be, as customary, parcelled out between the hydrostatic and viscous terms and cast as 
\begin{equation}
\bm{\nabla}\cdot\mathbb{P}=\bm{\nabla}P-\eta_s\nabla^2\vec{v}-\zeta\bm{\nabla}\left(\bm{\nabla}\cdot\vec{v}\right),
\end{equation}
with the pressure in the 2D Fermi-Dirac system \cite{Landau1976a,Giuliani2005a,Chaves2017}
\begin{equation}
P=\frac{\hbar v_F}{3 \pi}\big(\pi n\big)^{\frac{3}{2}}, \label{eq:FermiLiquidPress}
\end{equation}
and where 
the coefficients $\eta^s$ and  $\zeta$ represent the shear and bulk viscosity, respectively. Nonetheless, the bulk viscosity in graphene is extremely low \cite{Principi2016BulkSheet} and will, therefore, be neglected in this discussion.

Also, in the gradual channel approximation, the Fermi level varies slowly along the graphene, and so the field can be considered uniform and akin to a plane capacitor; with a capacitance per area $C\approx\varepsilon/d_0$ where $d_0$ is the graphene/gate separation and $\varepsilon$ the medium permittivity. Thus, the applied bias potential $U_\text{gate}$ is directly linked to the carrier density as $U_\text{gate}=en/C$ and so, the force term in \eqref{eq:model} is simply
\begin{equation}
   \bm{\nabla}\phi= \bm{\nabla}U_{\text{gate}}=\frac{ed_0}{\varepsilon}\bm{\nabla}n.\label{eq:Ugate}
\end{equation} 
Thus, since both the pressure and force terms are functions of the number density we group them as an effective pressure $\mathcal{P}$ such that 
\begin{equation}
    \bm\nabla\mathcal{P}=\bm\nabla P+\frac{e^2d_0n}{\varepsilon}\bm\nabla n
\end{equation}


\section{Boundary Layer Problem}

\begin{figure}[ht!]
    \centering
    \resizebox{0.8\linewidth}{!}{
    \begin{tikzpicture}[scale=0.6]
\filldraw[gray!25] (0,0) rectangle (10,-.3);
\draw[thin,-latex] (0,0) -- (10.3,0)node[anchor=north]{$x$};
\draw[thin,-latex] (0,0) -- (0,3.5) node[anchor=east]{$y$};
\draw[thick,densely dashed] (0,0) .. controls (1,1) and (4,2) ..  (10,2);

\draw[thin,latex-latex] (9,0.1)--(9,1.9);
\node[anchor=west] at (9,1) {$\delta(x)$};



\draw[very thin,-latex]  (-1,0.5)--(-.2,0.5);
\draw[very thin,-latex] (-1,1)--(-.2,1);
\draw[very thin,-latex]  (-1,1.5)--(-.2,1.5);
\draw[very thin,-latex]  (-1,2)--(-.2,2);

\node[] () at (-.6,2.5) {$U_0$};
\node[anchor=west] () at (2,2) {$U(x)$};
\node[anchor=west] () at (2,.8) {$u(x,y)\quad v(x,y)$};
\end{tikzpicture}
    }    
    \caption{Schematic representation of the 
    boundary layer after the injection of a flow $U_0$ near an interface. The flow is divided between two regions at a distance $\delta(x)$. Outside the boundary layer the flow $U(x)\rightarrow U_\infty$ is assumed to be unperturbed and parallel to the interface whereas inside the boundary layer the flow is described by $\vec{v}=u(x,y)\vec{\hat{x}}+v(x,y)\vec{\hat{y}}$}
    \label{fig:boundary_layer}
\end{figure}
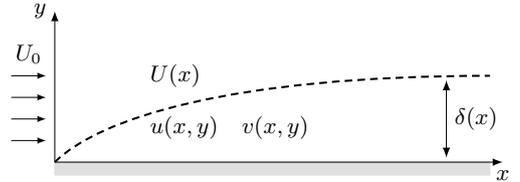

Although several works have been devoted to the viscous flow of electrons on graphene, akin to Poiseuille flow  \cite{Sulpizio2019,Torre2015a,Ku2020a}, for the flow to be in such conditions, of full development and entirely under the viscous effects, the longitudinal dimension of the flow must be significantly larger than the cross-section, so that it might be larger than the entrance length\cite{Schlichting2017Boundary-layerEdition}; that is, conform to the narrow channel approximation. In wider channels, or less viscous situations, we can expect that the flow is separated in two distinct regions, the uniform bulk and the boundary layer. Boundary layer is defined as the region of the flow where the velocity field is not uniform due to the influence of the viscous stress exerted on the fluid from the interface; it is bounded by a typical thickness $\delta$\cite{Schlichting2017Boundary-layerEdition}, as schematized on Fig. \ref{fig:boundary_layer}.   

\begin{figure}[ht!]
    \centering
    \includegraphics[width=\linewidth]{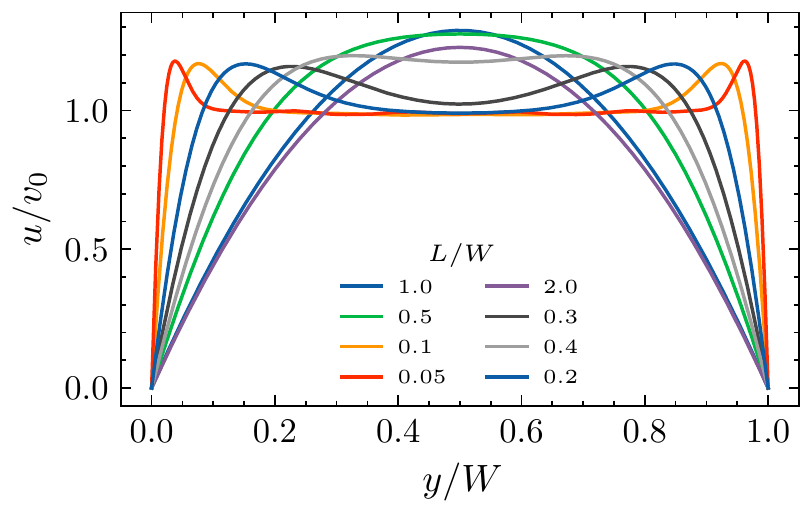}
    \caption{Velocity profile of viscous electronic flow ($\nu_s\equiv\eta_s/m^\star n=4/Lv_0$). Transition from Poiseuille to non-convex flow width the varying aspect ratio.}
    \label{fig:pouisseille_transition}
\end{figure}

While for incompressible fluids with a no-slip boundary condition Blasius theory \cite{Schlichting2017Boundary-layerEdition,Dewey1967} provides us a good description of the characteristic thickness of the boundary layer, the situation of Fermi liquid herein considered is more demanding. Not only is the fluid compressible due to the non-constant mass, but also the fluid at the edges may not be completely at rest and, on the contrary, slip along the boundary\cite{Kiselev2019BoundaryFlow}. 
Moreover, our simulations of the flow have showed that the velocity profile is not monotonic, displaying a maximum value near the edge before converging to the free flow limit, as can be seen in Fig.\ref{fig:pouisseille_transition}.  
This can be exploited to easily characterize the distinct regions of the flow, identifying the boundary layer thickness with the position of maximum velocity $\delta_{\rm max}$.

\subsection{Heuristic scaling law for $\delta_{\rm max}$}

\begin{figure}[h!]
    \centering
    \includegraphics[width=.85\linewidth]{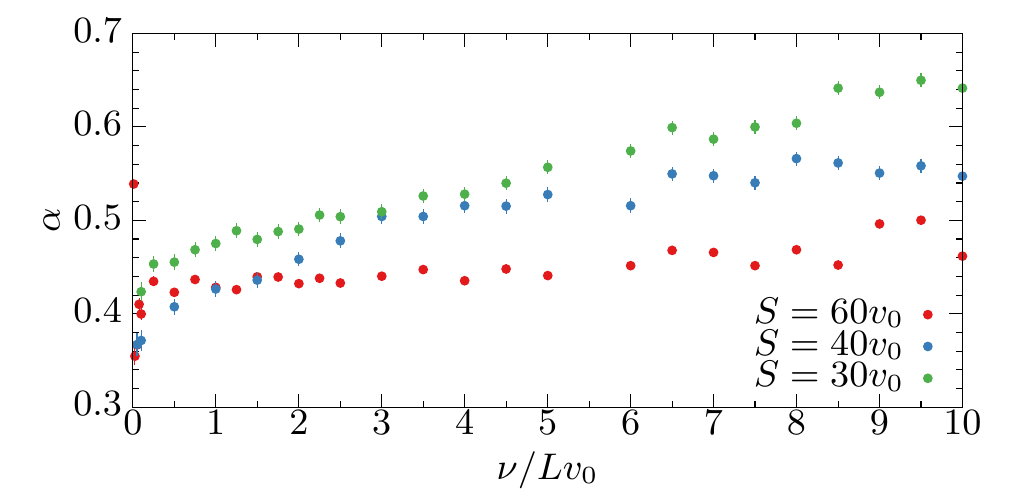}
    \caption{Fitted exponent parameter for $\delta_{\rm max}$ for varying kinematic viscosity and Mach number.}
    \label{fig:alpha_plot}
\end{figure}

From a dimensional analysis approach one can argue that the boundary layer thickness should follow a scaling law 
\begin{equation}
    \delta_{\rm max}\sim \frac{x}{{\rm Re}_x^\alpha},\label{eq:empirical_dmax}
\end{equation}
with ${\rm Re}_x$ the Reynolds number of the flow at position $x$ and $\alpha$ an exponent to be determined. Solving numerically the model \eqref{eq:model},  with no slip boundary condition and sufficiently wide aspect ratio, provides the velocity profiles to which \eqref{eq:empirical_dmax} can the fitted to retrieve $\alpha$. Our findings show that it increases marginally with kinematic viscosity and Mach number as patent on Fig. \ref{fig:alpha_plot}.

Although useful for empirical usage this numerical approach does not grant us full understanding of the velocity profile, thus, an analytical description of such profile was pursuit. 

\subsection{Analytical description of velocity profile}

In order to obtain a more in depth description of the flow we followed the usual  procedure to determine the laminar boundary layers equations \cite{Schlichting2017Boundary-layerEdition,Dewey1967}. From \eqref{eq:model} one gets, 
\begin{subequations}
\begin{equation}
    \frac{u}{2}\frac{\partial u}{\partial x}+v\frac{\partial u}{\partial  y}-\frac{u}{2}\frac{\partial v}{\partial y}-\frac{\eta_s}{nm^\star}\frac{\partial^2 u}{\partial y^2}+\frac{1}{nm^\star}\frac{\partial \mathcal{P}}{\partial x}=0,\label{eq:ble-xmoment}
\end{equation}
\begin{equation}
    \text{and}\quad\frac{1}{nm^\star}\frac{\partial \mathcal{P}}{\partial y}=0.\label{eq:ble-ymoment}
\end{equation}
\end{subequations}
Moreover, the introduction of a potential field $\psi$, such that 
$nu=\partial_y \psi$ and $nv=-\partial_x \psi$, automatically solves the steady-state of \eqref{eq:continuity}.

Equation \eqref{eq:ble-ymoment} imposes that the pressure and, consequently trough \eqref{eq:FermiLiquidPress}, the number density are solely functions of $x$. Thus, the  Dorodnitsyn-Howarth \cite{Schlichting2017Boundary-layerEdition} transformation for the $y$ coordinate yields 
\begin{equation}
    \eta = \frac{n(x)}{n_\infty}\sqrt{\frac{U_\infty}{2\nu x}}y
\end{equation}
which, along with the scaling of the potential field as
\begin{equation}
    \frac{\psi}{\sqrt{2\nu x U_\infty}n_\infty}= f(\eta),
\end{equation}
converts, after some algebra, Eq. \eqref{eq:ble-xmoment} to the nonlinear third-order equation 
\begin{equation}
    \frac{n^2}{n_\infty^2}f'''+ff''=\frac{x}{n}\frac{dn}{dx}\left({f'}^2+2\frac{S^2}{U_\infty^2}\right)\label{eq:BLE_F}
\end{equation}
with $S=\left(v_F^2/2+e^2 d v_F\sqrt{ n_0}/\varepsilon \hbar\sqrt{\pi}\right)^{1/2}$ the sound speed of the plasmonic fluid. This equation ought to be supplemented with the appropriate boundary conditions $f'(0)=0$ and $f'(\infty)=1$ representing the non-slip condition and asymptotic value of the velocity, respectively. Contrarily to regular boundary layers, Eq. \eqref{eq:BLE_F} can have overshoot solutions, i.e. with $f'>1$ in some region, correctly reproducing the behavior observed in the simulations. 





Let Eq. \eqref{eq:BLE_F} be recasted as 
\begin{equation}
     a f''' + f f' = \beta\left(f'^2+2m^2\right),
     \label{eq:BLE_Fp}
\end{equation}
with the obvious identifications, to ease its analysis. Note that this equation reduces to Blasius equation for $a=1$ and $\beta=0$.
All three parameters can, in first order, be considered constant as they only vary along the length of the channel. Regarding its values one expects, $a\simeq1$, as there should not occur a sizable variation of number density across the channel width. In the case of $\beta$, it encodes the density drop along the channel; thus, should be negative, as the density should not increase, but it is also forseeable that $|\beta|\ll1$. Finally, $m$, inverse Mach number, can be tuned between about 10 and 100, due to the properties of the graphene electron fluid in study\cite{Cosme2021}. 

In order to analyse \eqref{eq:BLE_Fp} and to obtain some insight on its solutions, the differential transformation method (DTM) \cite{Bervillier2012} can be performed, from which one obtains a recursive formula for the coefficients of the formal power series expansion of $ f(\eta) = \sum_{k=0}^{\infty} F_k\eta^k $ yielding
\begin{multline}
    F_{k+3} = -\frac{k!}{a(k+3)!} \Bigg[\sum_{q=0}^k (k-q+2)(k-q+1)F_{q-1}F_{k-q} \\
     -2\beta m^2-\beta\sum_{q=0}^k (q+1)(k-q+1)F_{q+1}F_{k-q+1}\Bigg].
\end{multline}
As the condition $f'(\infty)=1$ is incompatible with an expansion about the origin it was replaced by $f''(0) = \tau$, i.e. fixing the shear at the edge. Consequently, the first coefficients can be retrieved as
\begin{align*}
    F_0 &= 0 &, \quad F_2 &= \frac{\tau}{2} &, \quad F_4 &= \frac{m^2 \beta}{12a}  \\ 
    F_1 &= 0 &, \quad F_3 &=  \frac{m^2 \beta}{3a}   &, \quad F_5 &= -\frac{\tau^2}{120}+\frac{\beta(2m^2+\tau^2)}{60a}.
\end{align*}
It becomes clear that $m^2\beta$ is a figure of merit in the general behaviour of the solutions, setting them apart from the classical boundary layer profiles. 

\begin{figure*}
    \centering
    \includegraphics[width=0.3\linewidth]{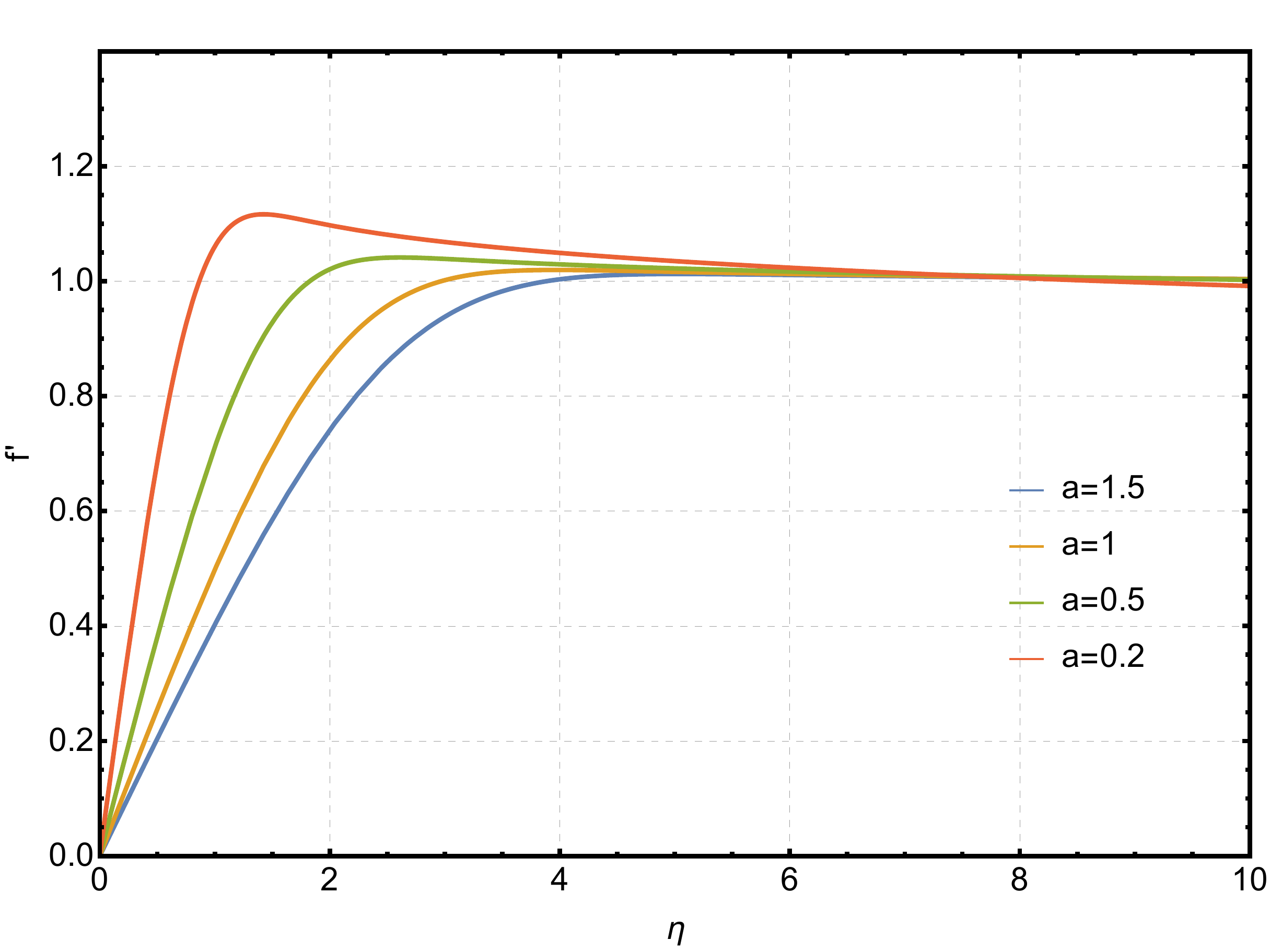}
    \hfill
    \includegraphics[width=0.3\linewidth]{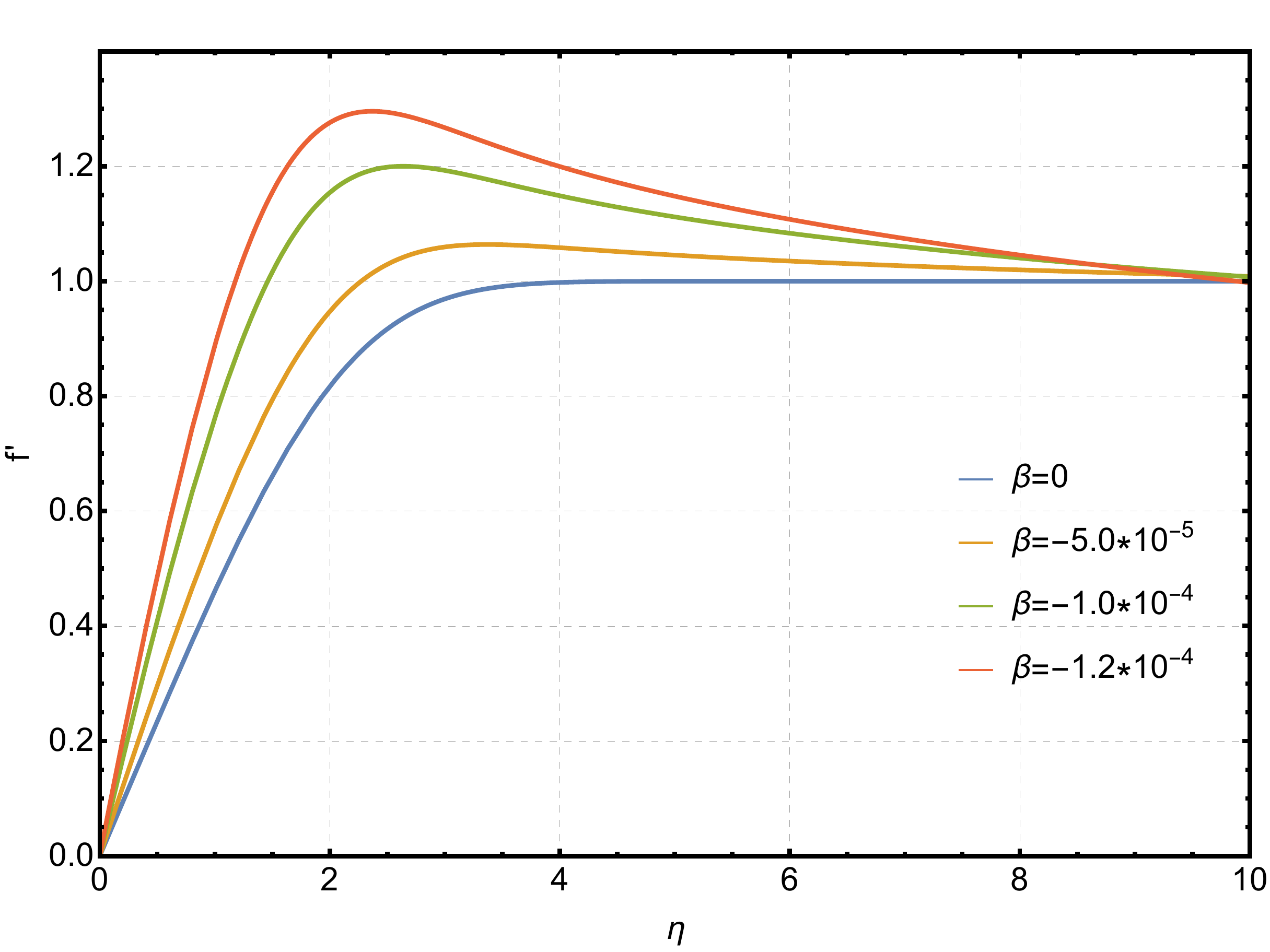}
    \hfill
    \includegraphics[width=0.3\linewidth]{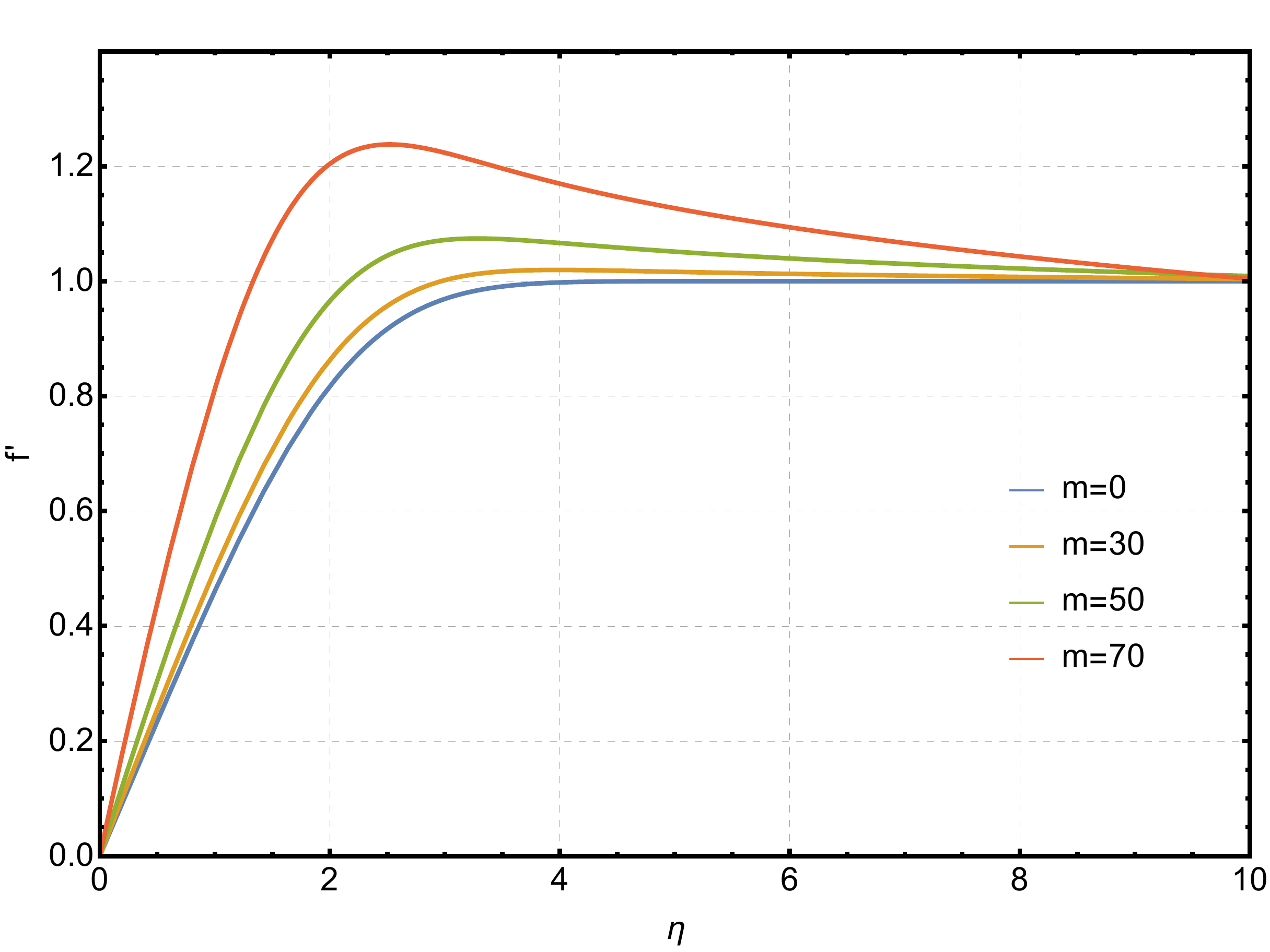}
    \vspace{-1 em}
    \caption{ Solutions of boundary layer equation $ a f''' + f f' = \beta\left(f'^2+2m^2\right)$. Sweeping parameters $a$ (left panel), $\beta$ (centre panel) and $m$ (right panel). Reference values of parameters when not being varied: $a=1$; $\beta=-2\!\times\!10^{-5}$; $m=30$.}
    \label{fig:variations}
\end{figure*}

Solving, numerically, the boundary layer equation \eqref{eq:BLE_Fp} we are able to recapture the characteristic non-monotonic behaviour observed in the fluid simulations. This aspect is clear in  Fig. \ref{fig:variations} where the influence of parameters variation can also be observed.

\begin{figure}
    \centering
    \includegraphics[width=\linewidth]{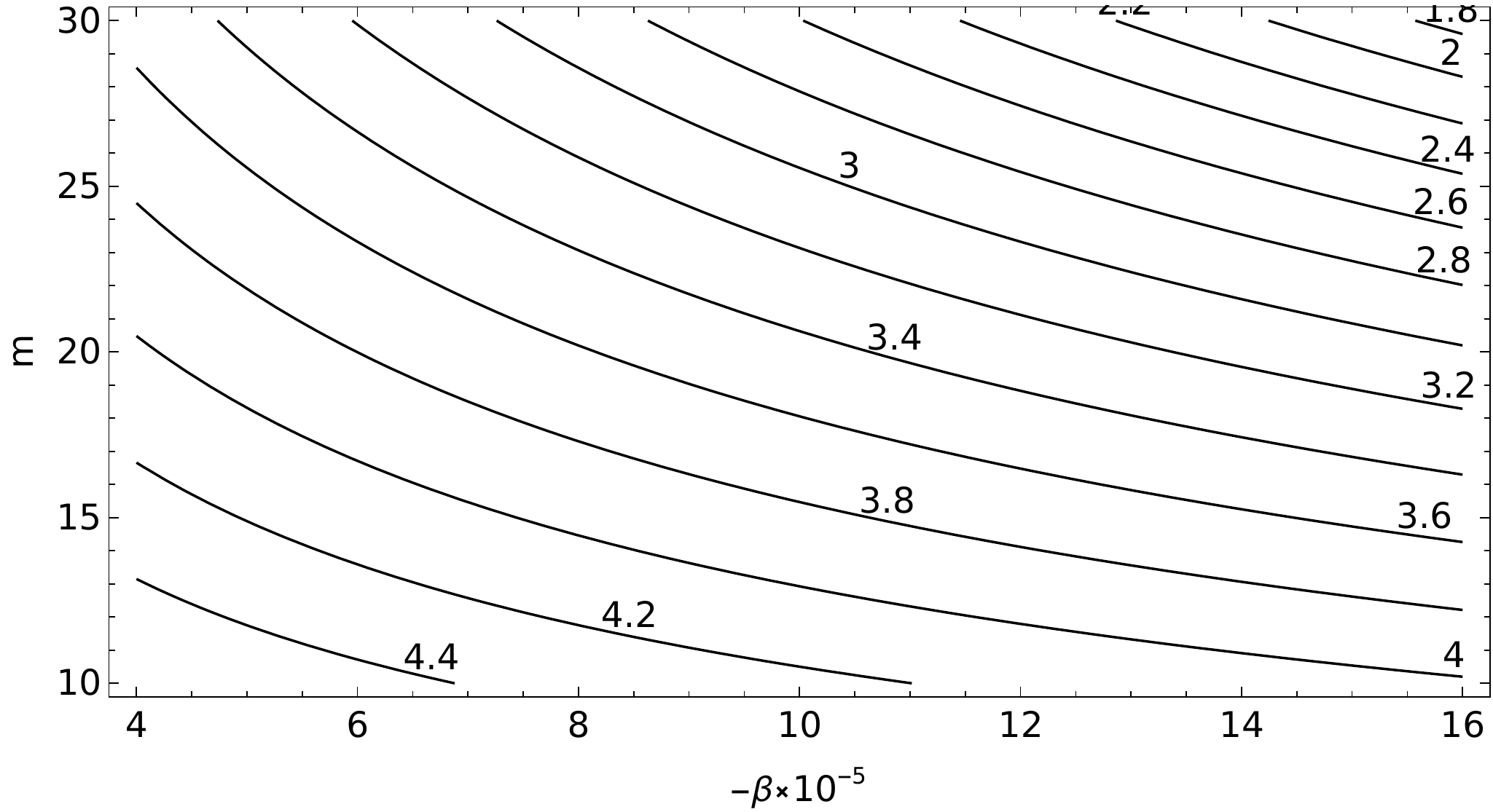}
    \caption{Argument of maximum of $f'$, $\eta_{\rm max}\propto\delta_{\rm max} $, as function of parameters $\beta$ and $m$.}
    \label{fig:eta_max}
\end{figure}

\section{Discussion}

Although not expected for regular fluids, the peaked profile of the velocity field is a prevailing feature of our simulations of graphene intermediate viscous regimes; signalling a transition from Ohmic to fully viscous flow.  Thus, while the non-monotonicity of $f'(\eta)$ impedes the usual technique of characterizing the boundary layer by the displacement or momentum thicknesses, $\delta_{\rm max}$ can be consistently used as $\delta(x)$ indicator. 
 
However, the simulation approach to determine $\delta_{\rm max}$ shows that de boundary layer lacks universality; since $\delta_{\rm max}$ power law has a small, yet noticeable, dependence on $S/v_0$ and $\nu$ itself. Nonetheless, we can estimate $0.4\lesssim\alpha\lesssim0.6$, in line with the expected square root signature. Furthermore, the functional dependence of $\delta_{\rm max}$ with flow parameters becomes evident with our model \eqref{eq:BLE_F}. Not only the Mach number appears explicitly, but also, the density drop $\beta$ must be dependent on the Mach and Reynolds numbers, albeit in a still unknown way.     

Regarding the generalized boundary layer model, we found good agreement between the solutions of \eqref{eq:BLE_Fp} and the simulated velocity profiles, particularly in replicating the velocity maximum. Showing that, even a minute pressure drop, around $\beta\approx-10^{-4}$, can be responsible for the non-convex flow profile.   

\vspace*{1.5cm}

\section{Conclusions}

In this work, we derive a generalization of boundary layer theory for viscous flow in a hydrodynamical description of the massless electrons in graphene. This model, characterized by the parameters $a=n/n_\infty$, $\beta=d\log n/d\log x$ and $m=S/U_\infty$, accurately predicts the non-monotonic velocity profile observed in our performed fluid simulations and the experimental observations reported in the literature. Our results, pointing towards the existence of a velocity peak near the edge, show that edge currents may show up in the system. This may lead to a better understanding of the topic of non-topological edge currents in graphene, which have been recently found in the literature \cite{Aharon-Steinberg2020}.

Moreover, our findings, both from the analytical model and the simulations, provide a valuable way to estimate the boundary layer thickness in experimental configurations as $\delta_{\rm max}\sim x/{\rm Re}_x^\alpha$ with $0.4 \lesssim \alpha \lesssim 0.6$. Such scaling law can be put in use when designing or analysing graphene circuitry.     

Looking towards the future, the effects of a slip length type boundary, i.e. where $f'(0)\neq0$, may introduce new velocity profile contours. We are confident that the model presented in this work is able to cope with such conditions and remain faithful to the experimental and numerical results; still, further investigation in such scenario is required. 

Finally, the inclusion of odd viscosity in situations with broken parity symmetry (namely with an applied magnetic field) may also be a feature of future interest.

\vspace*{4cm}

\begin{acknowledgments}

The authors acknowledge the funding provided by Fundação para a Ciência e a Tecnologia (FCT-Portugal) through Grant No. PD/BD/150415/2019 and through the exploratory project UTA-EXPL/NPN/0038/2019.

\end{acknowledgments}

\bibliographystyle{ieeetr}
\bibliography{references}

\end{document}